\documentclass{jfm}
\usepackage{graphicx}
\usepackage[section]{placeins}
\usepackage{float}
\usepackage{epstopdf, epsfig}
\usepackage[utf8]{inputenc}
\usepackage{textcomp}
\usepackage{multirow}
\usepackage{epstopdf}
\usepackage{pdfpages}
\usepackage{soul}
\usepackage{xcolor}
\usepackage{natbib}
\usepackage{subfigure}
\usepackage{hyperref}
\hypersetup{
	colorlinks = true,
	linkcolor  = blue,
	urlcolor   = blue,
	citecolor  = blue,
}

\shorttitle{Drag modification by dispersed droplets}
\shortauthor{C. Wang, L. Yi, L. Jiang and C. Sun}

\title{Turbulence drag modulation by dispersed droplets in Taylor-Couette flow: the effects of the dispersed phase viscosity}

\author{Cheng Wang\aff{1}, Lei Yi\aff{1} , Linfeng Jiang\aff{1}, \and Chao Sun\aff{1,}\aff{2}\corresp{\email{chaosun@tsinghua.edu.cn}}}

\affiliation{
	\aff{1}Center for Combustion Energy, Key Laboratory for Thermal Science and Power Engineering of Ministry of Education, Department of Energy and Power Engineering, Tsinghua University, Beijing 100084, China
	\aff{2}Department of Engineering Mechanics, School of Aerospace Engineering, Tsinghua University, Beijing 100084, China}

\begin{document}

\maketitle

\begin{abstract}
The dispersed phase in turbulence can vary from almost inviscid fluid to highly viscous fluid. By changing the viscosity of the dispersed droplet phase, we experimentally investigate how the deformability of dispersed droplets affects the global transport quantity of the turbulent emulsion.
Different kinds of silicone oil are employed to result in the viscosity ratio, $\zeta$, ranging from $0.53$ to $8.02$. The droplet volume fraction, $\phi$, is varied from 0\%  to 10\% with a spacing of 2\%.
The global transport quantity, quantified by the normalized friction coefficient $c_{f,\phi}/c_{f,\phi=0}$, shows a weak dependence on the turbulent intensity due to the vanishing finite-size effect of the droplets. The interesting fact is that, with increasing $\zeta$, the $c_{f,\phi}/c_{f,\phi=0}$ first increases and then saturates to a plateau value which is similar to that of the rigid particle suspension. By performing image analysis, this drag modification is interpreted from the aspect of droplet deformability, which is responsible for the breakup and coalescence effect of the droplets. The statistics of the droplet size distribution show that, with increasing $\zeta$, the stabilizing effect induced by interfacial tension comes to be substantial and the pure inertial breakup process becomes dominant. The measurement of the droplet distribution along the radial direction of the system shows a bulk-clustering effect, which can be attributed to the non-negligible coalescence effect of the droplet. It is found that the droplet coalescence effect could be suppressed as the $\zeta$ increases, thereby affecting the contribution of interfacial tension to the total stress, and accounting for the observed emulsion rheology. 

\end{abstract}

\section{Introdution}\label{sec:intro}
Droplets dispersed in flows - one of the omnipresent multiphase flows and also referred to as emulsions - involve complex interactions between phases and abundant dynamics of the dispersed phase. Despite their wide existence in industrial applications \citep{kilpatrick2012water, mcclements2007critical, spernath2006microemulsions}, the current emulsion studies, including but not limited to the macroscopic emulsion rheology and microscopic droplet statistics, are still limited, particularly for turbulent emulsions.

The rheology of emulsion is frequently characterized by the effective viscosity, $\mu_{e}$,  and in analogy with that of rigid particle suspension. The rheology of rigid particle suspension has been extensively discussed since the pioneering works of Einstein \citep{einstein1906, einstein1911}, which proposes a linear function of the effective viscosity concerning the volume fraction of the dispersed phase (i.e., $\phi$). Though this relation has been extrapolated to the semi-dilute regime by considering a higher-order term $O(\phi^2)$ \citep{batchelor1972determination, batchelor1972hydrodynamic}, empirical expression is still the only feasible method to predict $\mu_{e}$ at dense regime \citep{eilers1941viskositat, krieger1959mechanism, guazzelli2018rheology}. In addition, the previous relations are obtained based on the vanishing Reynolds number and may fail to capture the physics when the flow becomes turbulent. For example, it has been reported that, both in emulsions \citep{Pal2000ShearVB, yi2021global} and particle suspension \citep{adams2004influence, wang2022finite}, the effective viscosity of the system shows a shear-thinning/thickening behaviour in a turbulent flow. In fact, emulsions are significantly different from rigid particle suspensions. 
Many unique features of emulsions/droplets are absent in particle suspensions, such as the free-slip boundary condition, deformability, breakup, and coalescence. The effects of coalescence on the rheology of emulsions have been investigated in \cite{de2019effect} by introducing an Eulerian force to prevent droplet merging. In their work, the presence of coalescence could reduce the effective viscosity of the system by reducing the total surface of the dispersed phase, while the effective viscosity of the system is always greater than unity when the coalescence is inhibited. Since the total surface is proportional to the amount of energy transport \citep{crialesi2022modulation, schneiderbauer2022impact}, neglecting the droplet merging can lead to erroneous predictions \citep{de2019effect}. To bridge the gap between the droplet and rigid particles, many efforts have been devoted to revealing the physics of the rheology using such as viscoelastic sphere \citep{avazmohammadi2015rheology, rosti2018rheology, ye2019interplay} and deformable disk \citep{rosti2018suspensions, verhille2022deformability}. Overall, further studies are needed to explore the behaviours of emulsion rheology, especially in turbulent environments. 
 
The counterparts of turbulent emulsion rheology from the point of microscopic view are the mean droplet size, the droplet size distributions (DSDs), and the interactions between droplets, which highly depend on the flow configuration and the droplet property. One of the simplest cases is first addressed by \cite{kolmogorov1949breakage} and \cite{hinze1955fundamentals} characterizing droplets of low volume fraction dispersed in homogeneous and isotropic turbulence (HIT). By balancing the surface tension force and turbulent energy fluctuation, \cite{hinze1955fundamentals} proposes a critical Weber number $We_c\sim O(1)$ beyond which the droplet breakup happens, and further, the maximum droplet diameter that can stably exist (i.e., Hinze scale) is derived as
\begin{equation}\label{Hinze}
	d_H = \frac{We_c}{2}\left(\frac{\rho_c}{\sigma}\right)^{-3/5}\epsilon^{-2/5},
\end{equation}
where $\rho_c$ is the density of the continuous phase, $\sigma$ is the interfacial tension, and $\epsilon$ is the energy dissipation rate. The typical value of $We_c$ given by Hinze is 0.725. Studies have validated this relation in bubbly \citep{masuk2021simultaneous, chan2021turbulent} and emulsion \citep{mukherjee2019droplet, bakhuis2021,yi2021global} flows. However, the work of Hinze is based on the assumptions that the coalescence effects and the viscous dissipation are negligible, which are unfulfilled when the droplet volume fraction is high or the viscosity of both phases is non-negligible (as in the present work). \cite{roccon2017viscosity} performs simulations to investigate the effects of viscosity on droplet breakup and coalescence and shows that, like an increase of surface tension does, the breakup rate is suppressed with increasing the droplet viscosity, resulting in reduced droplet numbers and increased mean droplet size. In their work, \cite{roccon2017viscosity} shows that the Hinze relation is valid only if the droplets are less viscous than the continuous phase, i.e., when breakup essentially dominates the dynamics. Studies have also reported that the Hinze theory can be valid to some extent for specific flow configurations with inhomogeneity, such as the shear-homogeneous turbulence reported in \cite{rosti2019droplets}, where pure homogeneity is lost due to a net mean flow. Nevertheless, most correlations are obtained in isotropic turbulent conditions, while the role of turbulence inhomogeneity and anisotropy remains unknown \citep{crialesi2022modulation}. One example is that the values of $We_c$ could depart from the Hinze relation ( i.e., $We_c\sim O(1)$) and vary from 1 to 12, depending on the employed fluids and the flow configuration \citep{roccon2017viscosity}. More specifically, the presence of boundary layers could provide new mechanisms for droplet formation, rendering the theory obtained in HIT conditions unavailable. \cite{yi2022physical} finds that in Taylor-Couette (TC) flows, the Kolmogorov-Hinze theory obtained in HIT conditions fails to account for the formation of small droplets since the energy dissipation rate in the bulk is too weak. Using the Levich theory for inhomogeneous turbulent flow \citep{levich1962physicochemical}, they reveal that the droplet fragmentation occurs within the boundary layer and is controlled by the dynamic pressure caused by the gradient of the mean flow.

As a consequence of the breakup and coalescence, the DSDs is fundamental to the understanding of droplet dynamics. Recently, \cite{crialesi2022modulation} reports that the DSDs in HIT follow the $d^{-3/2}$ and $d^{-10/3}$ scaling, respectively, for droplets smaller and larger than $d_H$, which have also been found in previous studies of bubbles \citep{deane2002scale, riviere2021sub}.
While other previous numerical studies of emulsions could not resolve small droplets and study in detail interfacial dynamics since the use of diffuse-interface methods. Consequently, only $d^{-10/3}$ scaling for larger droplet are qualitatively reported \citep{mukherjee2019droplet, soligo2019breakage}. However, the DSDs could exhibit distinctive trends when the flow becomes anisotropy, corresponding to different mechanisms of droplet formation. In the work of \cite{yi2021global}, the experimental dataset of DSDs obtained in turbulent TC flow, where strong flow anisotropy exists, is well fitted by the Log-Normal distributions, suggesting that according to the authors the droplet formation is primarily controlled by fragmentation process. \cite{yi2021global} also finds that the DSDs could be fitted by a Gamma distribution \citep{villermaux2004ligament, villermaux2007fragmentation} but let it be an open question as to which fitting function is a better description of the DSDs. 
In general situations, the DSDs may not agree with any above-mentioned concise distributions when the droplet viscosity becomes the principal role. For example, \cite{de2017effect} finds that, by changing the droplet viscosity, the DSDs could transform from monomodal to bimodal distribution and broadens as the small droplets become smaller and the large ones larger. In their work, the DSDs in the case of high droplet viscosity are found to be fitted by a combination of two independent Gamma distributions, while few lights have been shed to uncover the underlying physical mechanism.

To summarize, in the most of previous studies, the rheology of turbulent emulsion is mainly discussed in the low-Reynolds-number regime and few have measured the global transport quantity. On the other hand, the investigation of droplet statistics is mainly limited to the negligible coalescence effect and HIT condition, which may lose their validity when the flow becomes anisotropy or the droplet property is changed.

The main goal of the work is to measure the effect of droplets on the global transport quantity of turbulence. In the present work, the droplet deformability (and the associated droplet coalescence and fragmentation) is altered by changing the viscosity of the droplet. By using the TC flow, the global transport quantity (torque) is accurately measured, allowing us to excavate the links between the global transport quantity and the local statistics of droplets (the DSDs and the droplet distribution along the radial direction of the system).

\begin{figure*}
	\centering
	\includegraphics[width=1\linewidth]{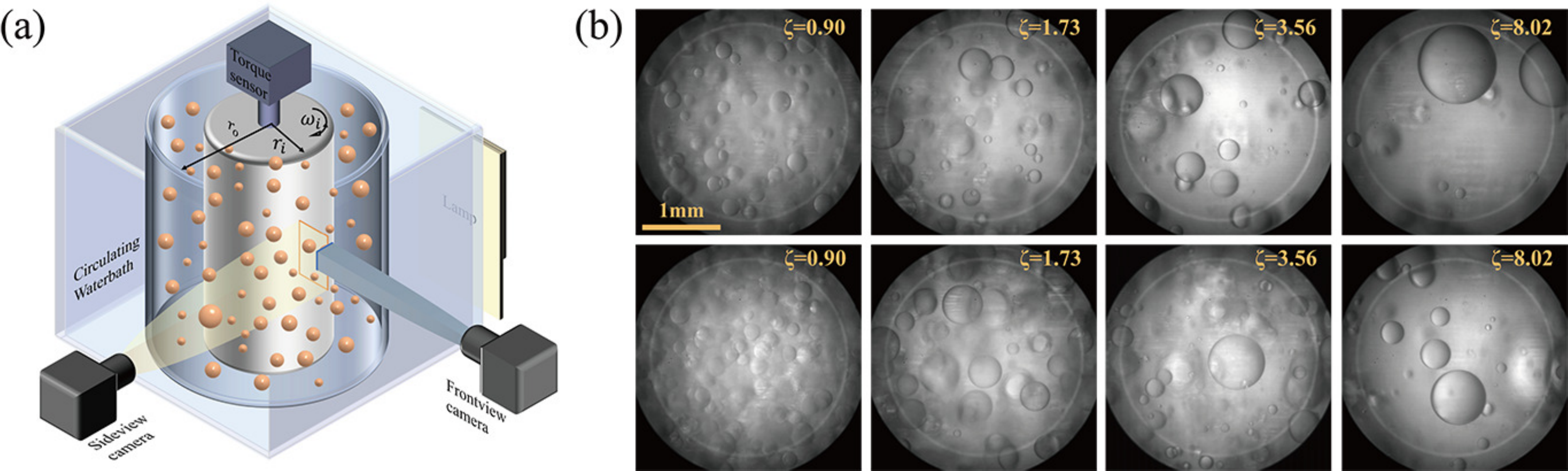}
	\vspace{-5 mm}
	\caption{(a) Sketch of the experiment setup. The droplet density is matched with the ethanol-aqueous solution by changing the volume fraction of ethanol of the ethanol-aqueous mixture. The images of the droplets between the gap are recorded from the side and the front view. The system temperature is maintained at $T=22\pm0.1$\ \textcelsius\ by a circulating bath. (b) Snapshots of the droplet morphology at $\dot{\gamma}=125$ s$^{-1}$. The value of $\zeta$ is depicted in each panel. Upper row for $\phi=2\%$ and lower row for $\phi=4\%$.}
	\label{fig1} 
\end{figure*}

\section{Experiment setup}\label{sec:exp}

The experiments are performed in a TC facility as shown in figure \ref{fig1}a, which has also been used in the previous work \citep{wang2022finite}. 
The radius of the inner and outer cylinder are $r_i$ = 25\ mm and $r_o$ = 35\ mm, and the height of the inner cylinder is $L$ = 75\ mm. A circulating bath maintains the system temperature at $T=22 \pm 0.1$\ \textcelsius. In experiments, the outer cylinder is fixed, while the inner cylinder is rotated. A torque sensor is used to measure the overall torque of the inner cylinder. The control parameter of TC flow is the shearing rate defined as
\begin{equation}\label{shear-rate}
\dot{\gamma} = \frac{\omega_i r_i}{r_o-r_i},
\end{equation}
where $\omega_i$ is the angular velocity of the inner cylinder. Alternatively, another  general parameter to quantify the turbulent intensity is the Reynolds number, $Re$, which for the TC flow can be defined as 
\begin{equation}\label{Re}
	Re=\frac{\rho_c (r_o-r_i)^2 \dot{\gamma}}{\mu_c},
\end{equation}
where $\mu_c$ is the dynamic viscosity of the continuous phase. Since the $Re$ depends on $\mu_c$, it varies when the viscosity of the dispersed droplet changes as we need to match the density. In this work, we mainly use the shearing rate, $\dot{\gamma}$, to quantity the turbulent intensity. The end-plate effect that comes from the secondary vortices generated at the end plates  \citep{bagnold1954experiments, hunt2002revisiting} has been considered, and the torque is calibrated using the same method in \cite{wang2022finite}. Here, only the torque due to the sidewall of the inner cylinder (TC flow), $\tau$, is used, which can be non-dimensionalized as
\begin{equation}\label{G}
	G=\frac{\tau}{2\pi L(\mu_c/\rho_c)^2},
\end{equation}
where $\rho_c$ is the density of the continuous phase. 

\begin{figure*}
	\centering
	\includegraphics[width=0.6\linewidth]{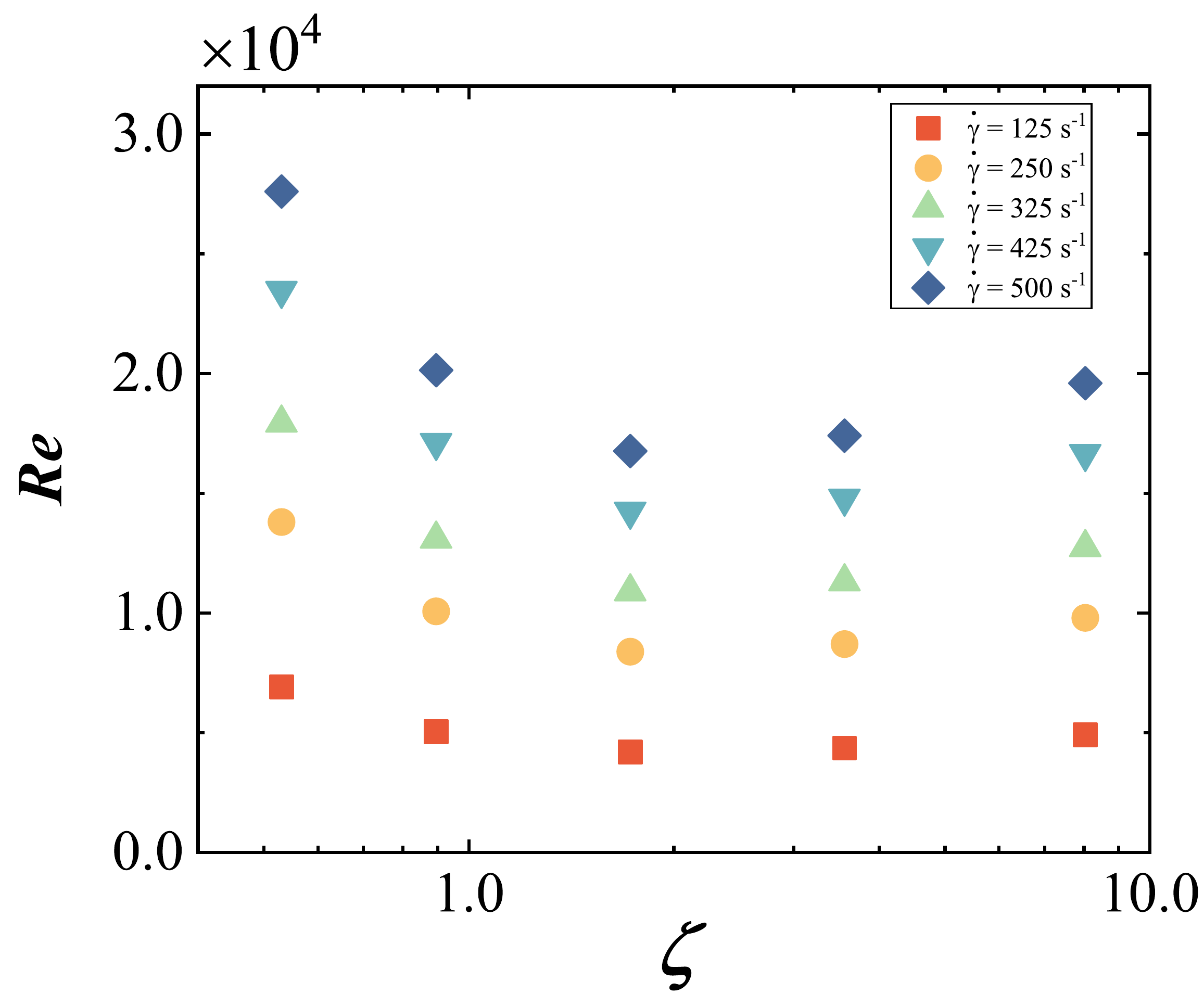}
	\vspace{-1 mm}
	\caption{The regimes of the control parameters investigated in the present work. For a fixed shearing rate $\dot{\gamma}$, the $Re$ is varied for different $\zeta$ due to the different concentrations ratios of the ethanol-aqueous solution used to match the droplet density.}
	\label{fig2} 
\end{figure*}

Silicone oil with different viscosities is employed to study the effects of droplet viscosity on the turbulence drag modulation, yielding another dimensionless parameter, i.e., the viscosity ratio defined as
\begin{equation}\label{ViscosityRatio}
	\zeta = \frac{\mu_d}{\mu_c},
\end{equation}
where $\mu_d$ is the viscosity of the dispersed phase. As has been done in previous work \citep{wang2022finite}, the torque measurements are performed for each volume-fraction of droplets, $\phi$, from 0\% to 10\% with a spacing of 2\%, so that to compare the different effects of rigid particles and droplets on the turbulence modulation. The details of the parameters of both phases are summarized in Table \ref{table1}. The viscosities of both phases are measured using a rheometer at $T=22$\ \textcelsius, while the density of the continuous phase (ethanol-aqueous solution) used to match the droplet density is interpolated with the data from \cite{khattab2012density} at $T=22$\ \textcelsius. In figure~\ref{fig2} we report the regime of the control parameters explored in the present work. Note that the silicone oil of low viscosity (Case-1 in this work) could slightly dissolve in ethanol. Therefore, for Case-1, we pre-dissolve the oil into the ethanol-aqueous solution until it is saturated, afterward, the exact amount of oil of the experiment cases is dispersed into the saturated oil-ethanol-aqueous solution. The deformability of droplets in turbulence can be connected to the Weber number and the Capillary number, which can be defined as
\begin{equation}\label{We}
	We = \frac{\rho_c u_b^2 d_p}{\sigma} = \frac{\rho_c d_p}{\sigma} \left(\frac{0.2(r_o^2-r_i^2)\dot{\gamma}}{r_i}\right)^2,
\end{equation}
\begin{equation}\label{Ca}
	Ca = \frac{\mu_c d_p \dot{\gamma}}{\sigma},
\end{equation}
where $\sigma$ is the interfacial tension, $d_p$ the mean droplet diameter, and $u_b\simeq0.2 \omega_i(r_i+r_o)$ the bulk velocity of the flow \citep{grossmann2016high,ezeta2018}.

\begin{table}
\begin{center}
    \resizebox{\linewidth}{!}{
	\begin{tabular}{cccccccc}
		\multicolumn{1}{c}{Case} & $\rho_d$ ($g/cm^3$)                  & $\mu_d$ ($mPa\cdot s$)                   & $\alpha$ & $\rho_c$ ($g/cm^3$)                  & $\mu_c$($mPa\cdot s$)                   & $\sigma$ ($mN\cdot m$) & $\zeta$ \\
        \\
		1                                         & 0.818                     & 0.79                       & 91.5\%                    & 0.817                     & 1.48                       & ——                         & 0.53                    \\
		2                                         & 0.873                     & 1.94                       & 75.0\%                    & 0.872                     & 2.16                       & 5.70                       & 0.90                    \\
		3                                         & 0.915                     & 4.71                       & 54.3\%                    & 0.916                     & 2.73                       & 6.25                       & 1.73                    \\
		4                                         & 0.935                     & 9.57                       & 44.4\%                    & 0.934                     & 2.69                       & 6.38                       & 3.56                    \\
		5                                         & 0.950                     & 19.5                       & 34.8\%                    & 0.951                     & 2.43                       & 8.67                       & 8.02 					  \\           
	\end{tabular}
    }
	\end{center}
	\caption{Parameters of the continuous and dispersed phases. $\rho$, $\mu$ are the density and viscosity with the subscripts $d$ and $c$ denoting the disperse and continuous phase, respectively, $\alpha$ is the volume fraction of ethanol in the ethanol-aqueous solution used to match the oil density, and $\sigma$ is the interfacial tension.}
	\label{table1}
\end{table}

In order to gain insights into the droplet statistics, two cameras are used to record the statistics of dispersed droplets from the front and the side views. Limited by the high volume-fraction of droplets, the droplet size and the droplet distribution along the radial direction of the system are manually detected. The details of the method of image analysis are referred to the previous works 
\citep{yi2021global,wang2022finite}. The image analysis, including the droplet morphologies in figure~\ref{fig1}b and the distributions in figures~(\ref{fig7}, \ref{fig8}) are only carried out at $\phi=2\%$ and $4\%$ for Cases (2-5). While for the higher volume-fractions of droplets (here in this work $\phi = 6\%$, $8\%$, $10\%$), the droplets occupy the entire field of view and the system becomes opaque, consequently the droplets cannot be detected. For Case-1 ($\zeta = 0.53$), the droplets are too small to be captured. Figure \ref{fig1}b shows an example of the snapshots of droplet morphology for different $\zeta$ and $\phi$ in the system.

\section{Results}\label{sec:results}
\subsection{The overall drag modification}\label{sec:drag}

\begin{figure*}
	\centering
	\includegraphics[width=1\linewidth]{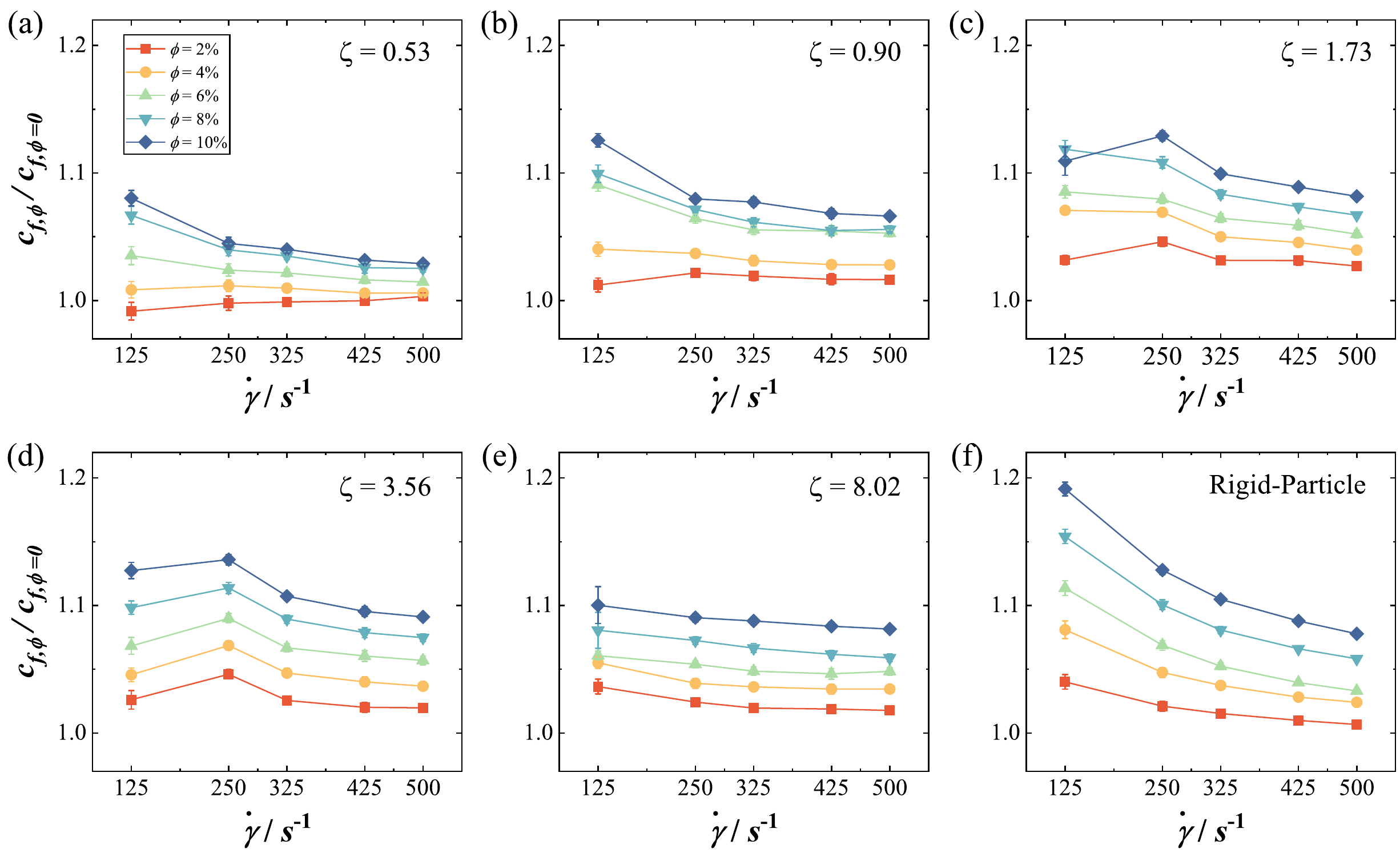}
	\vspace{-5 mm}
	\caption{The normalized friction coefficients of the TC system, $c_{f,\phi}/c_{f,\phi=0}$, for different $\zeta$. The value of $\zeta$ is shown in each panel, and panel (f) is the data for rigid spherical particles taken from \cite{wang2022finite}.	The accuracy of the experiments is indicated by the error bars, which are typically less than 1\%.}
	\label{fig3} 
\end{figure*}

In this section, we study the overall drag modification by changing the turbulent intensity and the viscosity of the dispersed droplet phase. The skin friction drag of the TC system is evaluated by the friction coefficient, $c_f$  \citep{van2011torque,wang2022finite}, which is defined as
\begin{equation}\label{cf}
	c_{f,\phi} = \frac{(1-r_i/r_o)^2}{\pi(r_i/r_o)^2}\frac{G}{Re^2}.
\end{equation}
Based on the definitions, it can be noticed that $G\sim \mu^{-2}$ and $Re\sim \mu^{-1}$. Therefore, the friction coefficient defined as $c_f\sim G/Re^2$ is not explicitly dependent on the viscosity, but only implicitly depends on viscosity through the torque embedded in $G$. The friction coefficient will not be affected if the flow properties (such as $Re$ and $G$) are computed using a general form of viscosity (e.g., $\mu = (1-\phi)\mu_c+\phi\mu_d$). To emphasize the magnitude of the drag modification induced by dispersed droplets, the friction coefficient of emulsions is normalized by that of the corresponding single-phase flow, i.e., the ethanol-aqueous solution. The normalized friction coefficient, $c_{f,\phi}/c_{f,\phi=0}$, is shown in figure~\ref{fig3}.

\begin{figure*}
	\centering
	\includegraphics[width=1\linewidth]{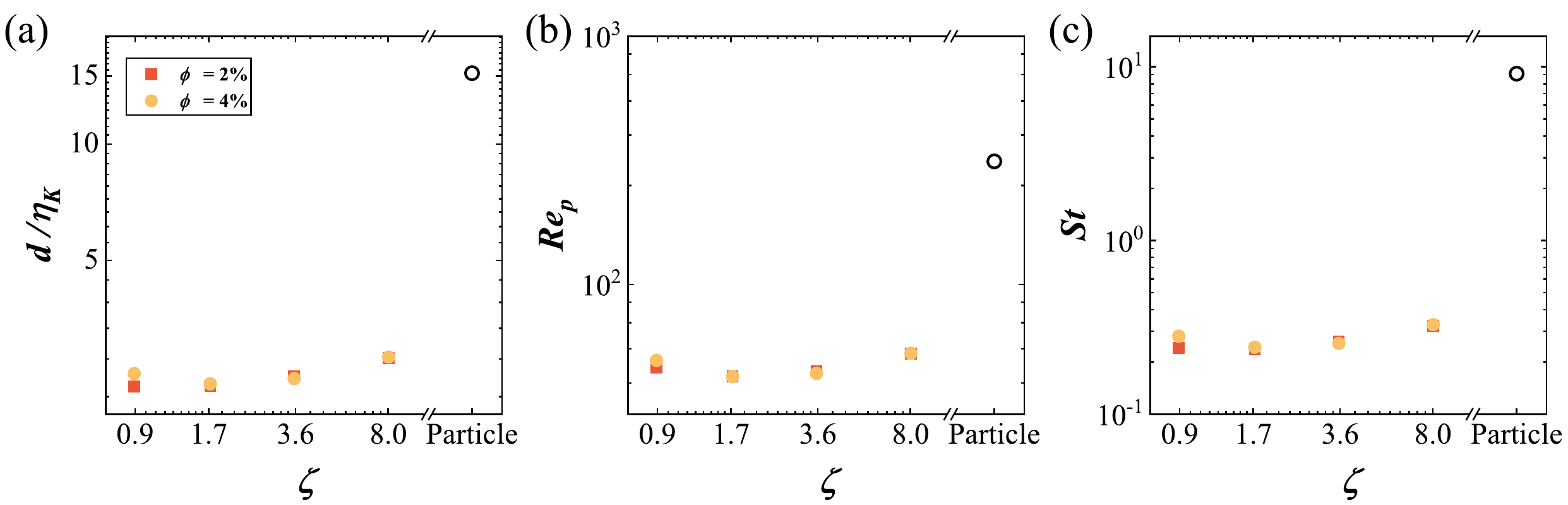}
	\vspace{-5 mm}
	\caption{(a) The dependence of (a) $d_p/\eta_K$, (b) $Re_p$, and (c) $St$ on the $\zeta$ at $\dot{\gamma}=125$ s$^{-1}$. The results of rigid particles (black hollow symbols) taken from \cite{wang2022finite} are also shown for comparison.} 
	\label{fig4} 
\end{figure*}

We first look into the effect of turbulent intensity, $\dot{\gamma}$ (or, $Re$), on the drag modulation, $c_{f,\phi}/c_{f,\phi=0}$, induced by droplet with fixed $\zeta$.
As shown in figure~\ref{fig3}(a,b), the $c_{f,\phi}/c_{f,\phi=0}$ of droplets with $\zeta<1$ monotonously decreases with increase $\dot{\gamma}$ (or, $Re$) in a general way, and shows a weak $\dot{\gamma}$ (or, $Re$) -dependence when compared to the rigid particles shown in figure~\ref{fig3}f. As the droplet viscosity increases to $\zeta>1$ (figure~\ref{fig3}(c,d)), it can be seen that the $c_{f,\phi}/c_{f,\phi=0}$ is somehow damped for the case of smallest $\dot{\gamma}$ (or, $Re$). This could be the result of the under-developed turbulent flow in the system at this moderate turbulent intensity \citep{grossmann2016high, wang2022finite}. For the rest of turbulent intensity (i.e., higher $\dot{\gamma}$ or $Re$) in figure~\ref{fig3}(c,d) , the $c_{f,\phi}/c_{f,\phi=0}$ monotonously decreases with increase $\dot{\gamma}$ (or, $Re$), but still shows weaker trends than the case of rigid particle. For even higher $\zeta$ (figure~\ref{fig3}(e)), the $c_{f,\phi}/c_{f,\phi=0}$ shows plateau for various $\dot{\gamma}$ (or, $Re$) and seems to depend only on the volume-fraction of the droplets. This may be due to the turbulence suppression caused by highly viscous dispersed droplets. In this case, the magnitude of the drag modulation could be mainly affected by the contribution of the surface tension, which is proportional to the volume fraction of the droplets\citep{de2019effect}.
The weak $\dot{\gamma}$ (or, $Re$) -dependence could be interpreted by looking into the finite-size effect of the droplet, which can be connected to the particle Reynolds number, $Re_p$, and Stokes number, $St$, 
\begin{equation}\label{Rep}
	Re_p = \frac{\rho_c u_bd_p}{\mu_c} = \frac{\rho_c d_p}{\mu_c}\frac{0.2(r_o^2-r_i^2)\dot{\gamma}}{r_i},
\end{equation}
\begin{equation}\label{St}
	St={\frac{\rho_pd_p^2}{18\rho_f[(r_o^2-r_i^2)r_i(r_o-r_i)]^{1/2}}(GRe)^{1/2}}.
\end{equation}
Figure~\ref{fig4} presents the $Re_p$ and $St$ along with $d_p$ normalized by the dissipation length scale, $\eta_K = (\nu^3/\epsilon)^{1/4}$, where $\nu$ is the kinematic viscosity of the continuous phase, and $\epsilon$ is the total energy dissipation rate of the flow defined as $\epsilon = \tau\omega/(\pi\rho L(r_o^2-r_i^2))$ \citep{van2007bubbly}. For the droplets, one can see that $d_p/\eta_K\simeq2$ for various $\zeta$, resulting in both the $Re_p$ and $St$ much smaller than that of the rigid finite-size particle in our previous study. Given that the $Re_p$ and $St$ in this work are around $50$ and $0.3$ (see figure~\ref{fig4}(b, c), respectively), the droplets behave like fluid tracers \citep{toschi2009lagrangian,grossmann2016high,mathai2020bubbly,brandt2022particle}, and modify the turbulence mainly by altering its volume-fraction \citep{voth2017anisotropic}, accounting for the $\phi$-dependence of $c_{f,\phi}/c_{f,\phi=0}$. As the $Re$ increases, the droplet size decreases in $d_p\sim Re^{-6/5}$ \citep{yi2021global}, giving $Re_p\sim Re^{-0.2}$ and $St\sim Re^{-1.9}$. Therefore, the finite-size effect of droplets becomes negligible at high $Re$ in the current turbulent TC system, in accordance with the weak $Re$-dependence of the $c_{f,\phi}/c_{f,\phi=0}$ for various $\zeta$.

\begin{figure*}
	\centering
	\includegraphics[width=1\linewidth]{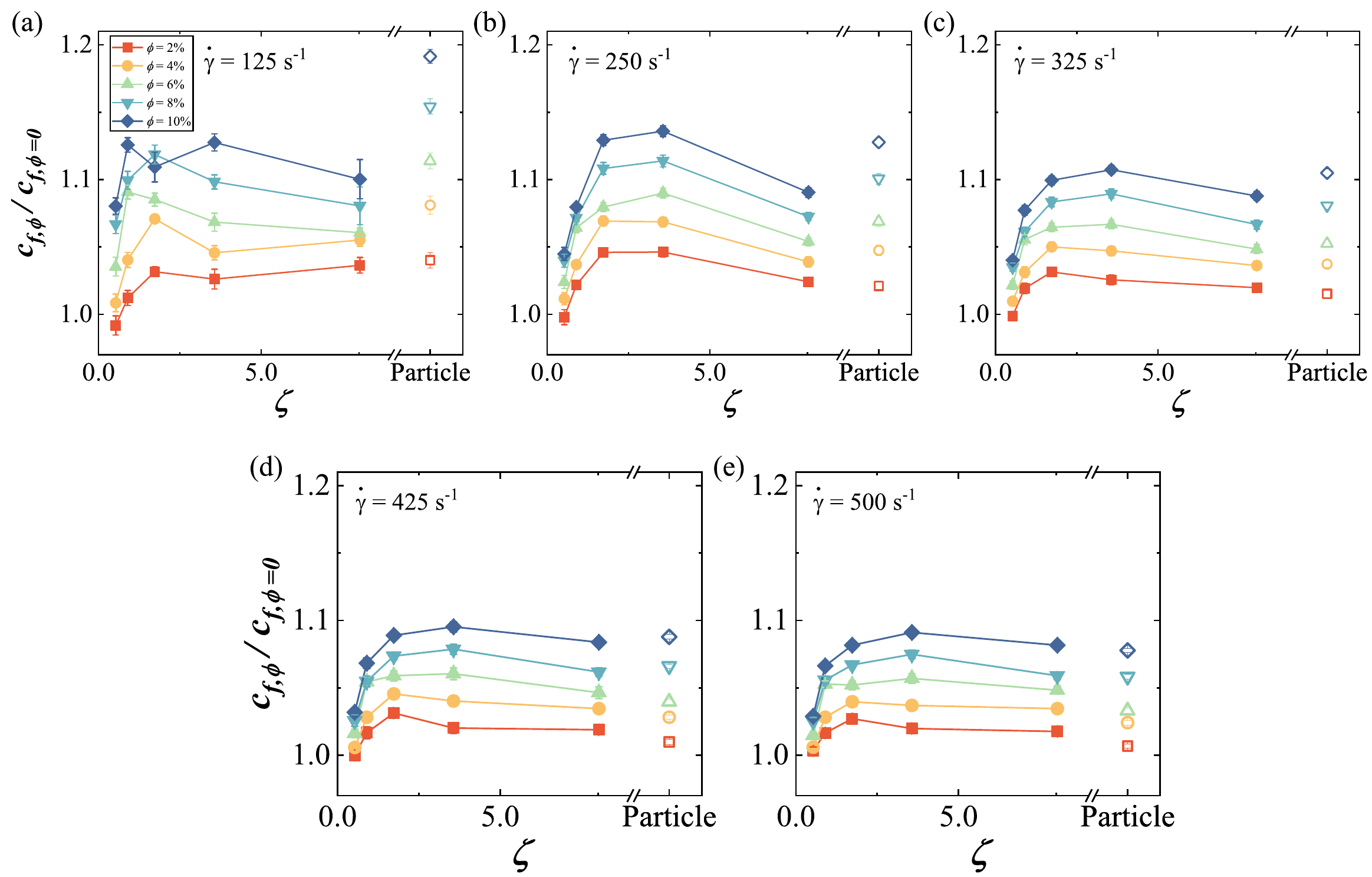}
	\vspace{-5 mm}
	\caption{The dependence of $c_{f,\phi}/c_{f,\phi=0}$ on the $\zeta$ for different shearing rate. The value of $\dot{\gamma}$ is shown in each panel. The results of rigid particles (hollow symbols) taken from \cite{wang2022finite} are also shown for comparison.} 
	\label{fig5} 
\end{figure*}

It is worth mentioning that for different $\zeta$, whether it is smaller than or larger than 1, the $c_{f,\phi}/c_{f,\phi=0}$ is always larger than unity, indicating that in the current system dispersed oil droplet always increases the overall drag of the system regardless their viscosity. This persistent drag increment may be counter-intuitive, as one might expect the dispersed phase with low viscosity to induce a drag reduction by reducing the effective viscosity of the system. However, it has been shown in \cite{de2019effect} that by introducing the extra interfacial tension and the viscous stress, the presence of droplets could provide a comparable, or even larger, contribution to the total stress than the reduced stress in the continuous phase. As for the dispersed droplets with high viscosity, the overall drag of the system is enhanced though the turbulent intensity might be suppressed. This could be understood since the turbulence attenuation is not necessary to connect to an overall drag reduction and vice versa, as suggested in our previous work \citep{wang2022finite}. Here we note that the stress increments of turbulent emulsions mainly come from the interfacial tension and the viscous stress of the dispersed droplets, whereas the one for the rigid particle suspension is the frictional contact between solid surfaces to compensate for the reduction in fluid turbulent stress. Therefore, for turbulent emulsions in TC turbulence within the current parameter regimes, the global transport efficiency, i.e., the overall drag of the system, is always increased by introducing dispersed droplets into the system. Future studies are needed to explore in broader regimes the behaviour of the drag modification induced by dispersed droplets.

We next inspect how the drag modulation induced by droplets changes as the $\zeta$ increases for fixed turbulent intensity. In figure \ref{fig5} we show the information of figure \ref{fig3} in the form of $c_{f,\phi}/c_{f,\phi=0} \sim \zeta$. For a fixed $\dot{\gamma}$,  $c_{f,\phi}/c_{f,\phi=0}$ first increases with increasing $\zeta$, which could be attributed to the attenuated coalescence at higher $\zeta$, and consequently results in an increment of surface tension contribution, and therefore, a higher effective viscosity \citep{de2019effect}. However, once the $\zeta$ exceeds a critical value (in the current work, $1.7\leq\zeta_c\leq3.8$), the $c_{f,\phi}/c_{f,\phi=0}$ of droplet approaches a plateau value. One could notice that this plateau value of $c_{f,\phi}/c_{f,\phi=0}$ for droplet is smaller than that for the case with rigid finite-size particles at small $Re$ or  $\dot{\gamma}$ (figure \ref{fig5}a), while when the flow is highly turbulent ($Re\sim O(10^4)$), as shown in figure \ref{fig5}(c-e), the difference between the turbulent emulsion and the particle suspension becomes negligible. This could be due to the decreasing significance of the contribution of the particulate phase to the total stress at high $Re$ \citep{wang2022finite}. Note that, when the $Re$ or $\dot{\gamma}$ is relatively small, the turbulence may be suppressed with dispersing droplets of high viscosity \citep{crialesi2022modulation}, and thus, accounting for the drag descent occurs at moderate $\zeta$ (figure~\ref{fig5}(a,b)). Nevertheless, the similarity between the drag of emulsions and rigid particle suspensions increases with increasing $\zeta$, which could be, as shown in later sections, due to the suppressed droplet deformability at higher $\zeta$. Based on the current results, one reasonable speculation could be that the difference in the global transport (drag) of turbulent emulsion system and rigid particle suspension would become even smaller with further increasing the $\zeta$ in turbulent Taylor-Couette system. Of course, more studies in wider parameter regimes will be needed to draw a concrete conclusion on this issue. 

\subsection{Droplet size distribution}\label{sec:size}
As discussed above, the drag of the turbulent emulsion shows a $\zeta$-dependence which could be connected to the droplet deformability. To dive into the $\zeta$-dependence of droplet microstructure and its relationship with the $c_{f,\phi}/c_{f,\phi=0}$, we look into the DSDs obtained from image analysis. Here we perform image analysis at the smallest $Re$ (or $\dot{\gamma}$) for $\phi =2\%$ and $4\%$. One may have noticed that the data quality of drag modulation is better when the flow is highly turbulent (e.g, figures~\ref{fig5}(d,e)) as the turbulence may be not well-developed at small $Re$ (or $\dot{\gamma}$). Intuitively, performing image analysis at high  $Re$ (or $\dot{\gamma}$) could be better to reveal the physics, which unfortunately can not be attained in the current work. Since the droplet size decreases with respect to $Re$ in $d_p\sim Re^{-6/5}$ \citep{yi2021global}, it means that at high $Re$ (or $\dot{\gamma}$) the droplet borders will soon become undetectable to the cameras. In the current experiments, at the smallest $Re$ (or $\dot{\gamma}$) the droplet borders can be well captured, while at higher $Re$ (or $\dot{\gamma}$), the droplet morphology from the front view becomes unclear and the droplet locations can not be distinguished from the side view. Additionally, the trendings of drag modulation at the smallest $Re$ (or $\dot{\gamma}$) and $\phi =2\%, 4\%$ (figure~\ref{fig5}a) show a similar behavior to that of higher $Re$ (or $\dot{\gamma}$), i.e, a two-stages evolution process. Therefore, we also note that the image analysis obtained at small $Re$ (or $\dot{\gamma}$) could provide valuable information for the understanding of the flow at higher $Re$ (or $\dot{\gamma}$). The relatively poor data quality of drag modulation at higher $\phi$ in figure~\ref{fig5}a could be attributed to the turbulence suppression induced by the highly viscous dispersed phase of high volume fraction. 

\begin{figure*}
	\centering
	\includegraphics[width=1\linewidth]{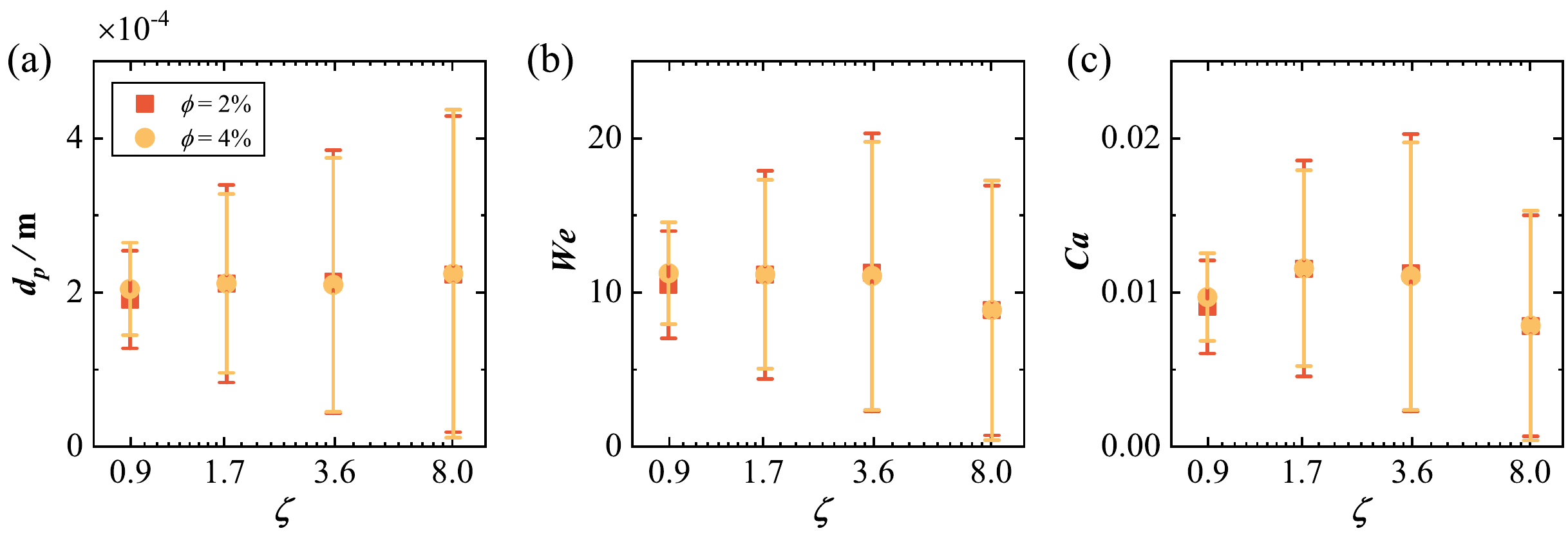}
	\vspace{-5 mm}
	\caption{The dependence of (a) the mean droplet diameter and (b) $We$ number (c) $Ca$ number on the $\zeta$ at $\dot{\gamma}=125$ s$^{-1}$.
    The error bars in (a) indicate the interval of standard deviation of the droplet size distribution (i.e. $d_p-\sigma \leq d \leq d_p+\sigma$). The upper (lower) bound of the error bars shown in (b,c) indicate the value of $We$ and $Ca$ corresponding to the upper (lower) bound of the droplet size shown in (a).
    } 
	\label{fig6} 
\end{figure*}

Figure~\ref{fig6}a shows that the mean droplet diameter weakly depends on the $\zeta$ and $\phi$, while the size range strongly increases as the $\zeta$ increases. Based on the mean droplet diameter and its standard deviations, the $We$ and $Ca$ defined by equations~(\ref{We},\ref{Ca}) are obtained, and the results are shown in figure~\ref{fig6}(b,c). In principle, the $We$ and $Ca$ decrease with increasing $\zeta$, suggesting that the deformability is suppressed for droplets with high viscosity. As one can see from figure~\ref{fig6}a, as the $\zeta$ increases, the maximum droplet size increases, and the probability of finding large droplets becomes higher, which is confirmed by the snapshots presented in figure~\ref{fig1}b and in accordance with the conclusions in \cite{crialesi2022modulation}.

\begin{figure*}
	\centering
	\includegraphics[width=1\linewidth]{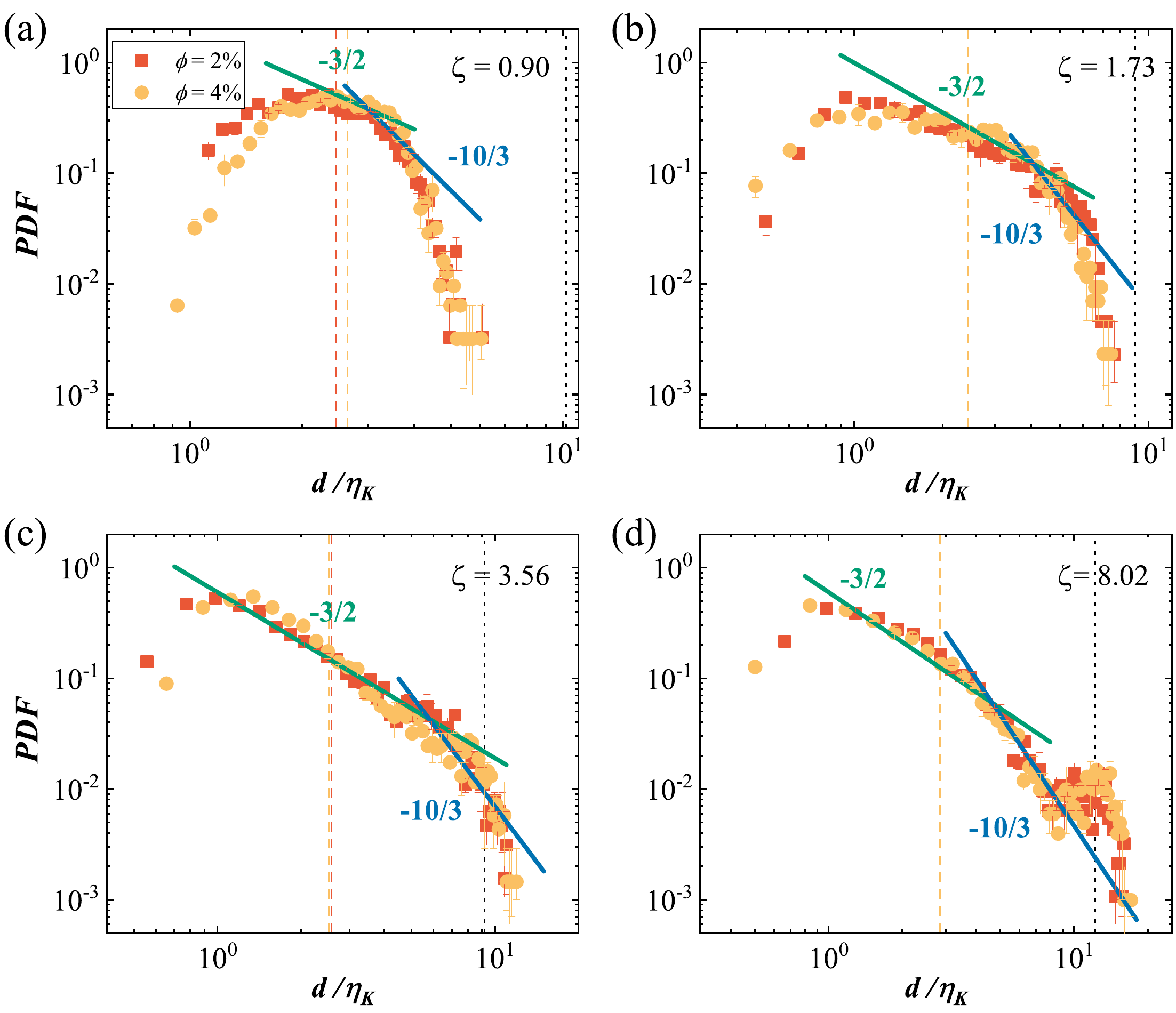}
	\vspace{-5 mm}
	\caption{The DSDs for different viscosity ratio at $\dot{\gamma} = 125$ s$^{-1}$. The value of $\zeta$ is shown in each panel. The droplet sizes are normalized by the dissipation length scale $\eta_K$ of each single-phase flow. The coloured vertical dashed lines denote the mean droplet size for $\phi=2\%$ and $\phi=4\%$, respectively (same colour code as the DSDs). The black vertical dotted line indicates the Hinze scale for each case.  The statistical errors are indicated by the error bars. The scaling relation $d^{-3/2}$ (green) and $d^{-10/3}$ (blue) are shown for comparison. Note that the expressions of the scaling relation are different in each panel.
	} 
	\label{fig7} 
\end{figure*}

To find out how the viscosity ratio affects the droplet breakup and coalescence, in figure~\ref{fig7} we also show the best fitted $d^{-3/2}$ and $d^{-10/3}$ scaling relations of each panel next to the experiment data. It has been reported that, in both the bubbly flow \citep{riviere2021sub} and the emulsion flow \citep{crialesi2022modulation} in HIT conditions, the DSDs shows a $d^{-3/2}$ scaling when $d<d_H$, which could be due to the increasing importance of surface tension effect for small bubble/droplet \citep{deane2002scale}; while when $d>d_H$, the DSDs are well described by the $d^{-10/3}$ scaling, which is first proposed by \cite{garrett2000connection} and later verified by \cite{deane2002scale}. Here $d_H$ is the Hinze scale defined as the maximum droplet diameter that can stably exist in turbulent emulsions.
The $d^{-10/3}$ scaling could be derived with the assumption that the large bubble ($d>d_H$) experiences a purely initial breakup process at a rate depending on the dissipation rate $\epsilon$ and the rate of supply of the dispersed phase.

As shown in figure~\ref{fig7}a, the DSDs rapidly deviate from the $d^{-10/3}$ scaling at the large size range due to the non-negligible coalescence effects at this small $\zeta$.
For small droplet size, the $d^{-3/2}$ scaling could only capture the distribution to some extent in a narrow range, hinting that the interfacial tension effect might be activated for only a limited population of small droplets; while the DSDs rapidly fall off when the droplet size further decreases, which might be due to the finite precision of image analysis and failing to capture the droplets with extremely small sizes. This could be the reason that the DSDs shown in figure~\ref{fig7}a agree with \cite{mukherjee2019droplet} in the range of small sizes. In the studies of \cite{mukherjee2019droplet}, $d^{-10/3}$ scaling for large droplets ($d>d_H$) is reported, while the droplets of small size (($d<d_H$)) are prone to dissolution and could not be resolve due to the limitation of the employment of diffuse-interface methods. As suggested by the simulation results from \cite{crialesi2022modulation}, the $d^{-3/2}$ scaling might still be valid for the significantly small droplets, which unfortunately could not be resolved in the current experiments. 

On the other hand, with increasing $\zeta$, the experimental data exhibits stronger coincidences with both two scalings. In particular, for the cases of the two largest viscosity ratios (figure~\ref{fig7}(c,d)), the DSDs could be generally described by these two scaling relations with a transition point at around $d/\eta_K \simeq 5\sim 6$. The pronounced consistency between the DSDs and the $d^{-10/3}$ scaling at larger $\zeta$ suggests that the assumption proposed by \cite{garrett2000connection} might be fulfilled in turbulent emulsion when the droplets are highly viscous, i.e., the large bubbles/droplets may experience a pure initial breakup process. Note that, at this relatively low shearing rate ($\dot{\gamma}=125$ s$^{-1}$), the droplets with high viscosity could be hard to be broken into pieces and may remain at the initial states, therefore accounting for the secondary peak appeared at the large size in figure~\ref{fig7}d. As for the small droplets at larger $\zeta$ (see figure~\ref{fig7}(c,d)), the DSDs principally follow the behaviour described by the $d^{-3/2}$ scaling. This could be understood considering that in the present work the droplet with large $\zeta$ possesses large surface tension, which could prevent the small droplets from merging into larger ones, thus maintaining the $d^{-3/2}$ scaling in this dynamic equilibrium system.

Given the definition of the Hinze scale ($d_H$) in equation~\ref{Hinze}, we can easily obtain the Hinze scale for various cases by taking the $\epsilon$ being the total energy dissipation rate of the flow as we mention above. As shown in figure~\ref{fig7}, unlike the cases in HIT conditions\citep{deane2002scale, mukherjee2019droplet, soligo2019breakage, riviere2021sub, crialesi2022modulation}, the scale of the transition point is not in line with $d_H$ (the black dotted line), and in fact are smaller than $d_H$. As suggested by \cite{mukherjee2019droplet} in HIT conditions, the length scale at which the DSDs transition into the $d^{-10/3}$ scaling could be predicted by that at which inertia and surface tension become comparable. Therefore, it sounds reasonable that this length scale of transition in HIT conditions can be approximated by the Hinze scale, which is defined by balancing the surface tension force and turbulent energy fluctuation.

However, the situation is different for TC turbulence where strong inhomogeneity and anisotropy exist. It has been recently reported that in TC flows \citep{yi2022physical}, the fluctuation in the bulk is too weak to generate droplets of the observed size. Instead, they proposed that the droplet fragmentation occurs within the boundary layer and is controlled by the dynamic pressure caused by the gradient of the mean flow, which can be accounted for using the Levich theory for inhomogeneous turbulent flow \citep{levich1962physicochemical}. This could partially be the reason that in the present work the scales of the transition point are smaller than the Hinze scale. For HIT flow, before it turns to $d^{-10/3}$ scaling, the $d^{-3/2}$ scaling for the small droplet is supposed to extend to the scale of $d_H$, where the surface tension can no longer sustain the large droplet under fragmentation of turbulent fluctuation. While for TC flow, droplet fragmentation could occur within the boundary layers, where the fluctuations are much stronger than in the bulk region. In other words, the scale that the surface tension can prevent the droplet from being fragmented by the turbulent fluctuation of the boundary layers is much smaller than the $d_H$. Therefore, the droplets in TC flow could become unstable at a smaller size than that in HIT flow, resulting in an earlier transition to $d^{-10/3}$ scaling before the scale of $d_H$. It should be noted that the experimental data reported here follows both scalings only in narrow ranges, typically less than one decade. As we mentioned above, the extension of $d^{-3/2}$ scaling to the extremely small size in this work is mainly limited by the experimental techniques to resolve extremely small droplets. While for the $d^{-10/3}$ scaling, its indistinct emergence could be attributed to the small size of the limited domain in the present work, which prohibits the emergence of large droplets. Additionally, the $Re$ in this work is still in a moderate regime, which might result in a relatively large dissipation length scale and thereby suppress the emergence of $d^{-10/3}$ scaling in the range of large scale. It is expected to further verify these two scalings in the future in broader ranges of size by improving the setup or increasing the turbulent intensity.

So far, one could curtly picture the physical process of the breakup and coalescence of droplets depending on the droplet deformability. (\romannumeral1) Droplets with small $\zeta$ possess higher deformability, promoting the large deformation to occur under the turbulence shearing. In this case, the coalescence effect is of comparable importance to the breakup effect and therefore is non-negligible. (\romannumeral2) Droplets with large $\zeta$ are hard to be sheared to deform by the turbulence. Once the large droplets exposed to the large shearing stress ($\tau \sim \rho {u^\prime}^2d$) are broken, part of the daughter droplets could be stabilized by the strong interfacial tension. This lack of unstable small droplets results in a reduction of the significantly small and large droplets, as confirmed by the ends of the DSDs shown in figure~\ref{fig7}(c,d). The shortage of large droplet formation indicates that the coalescence effect becomes negligible, or in other words, the large droplets experience an initial breakup process as assumed by \cite{garrett2000connection}.

At the end of this section, let us recur to the subject of uncovering the connection between DSDs and emulsion drag ($c_{f,\phi}/c_{f,\phi=0}$). The vanishing of the coalescence effect at higher $\zeta$, as demonstrated by the DSDs analysis, gives rise to an increment of the total surface area. Therefore, the increasing trend of $c_{f,\phi}/c_{f,\phi=0}$ concerning $\zeta$ results from the increment of the interfacial tension contribution to the total stress, as discussed in the previous section. When the interfacial tension effect becomes strong enough to maintain the droplet shape as the $\zeta$ further increases, the droplets are hard to deform and behave in a way like a rigid body, accounting for the approaching of $c_{f,\phi}/c_{f,\phi=0}$ of emulsion to the rigid particle suspension at large $\zeta$.

\subsection{Droplet distribution along the radial direction of the system}\label{sec:radial}

\begin{figure*}
	\centering
	\includegraphics[width=1\linewidth]{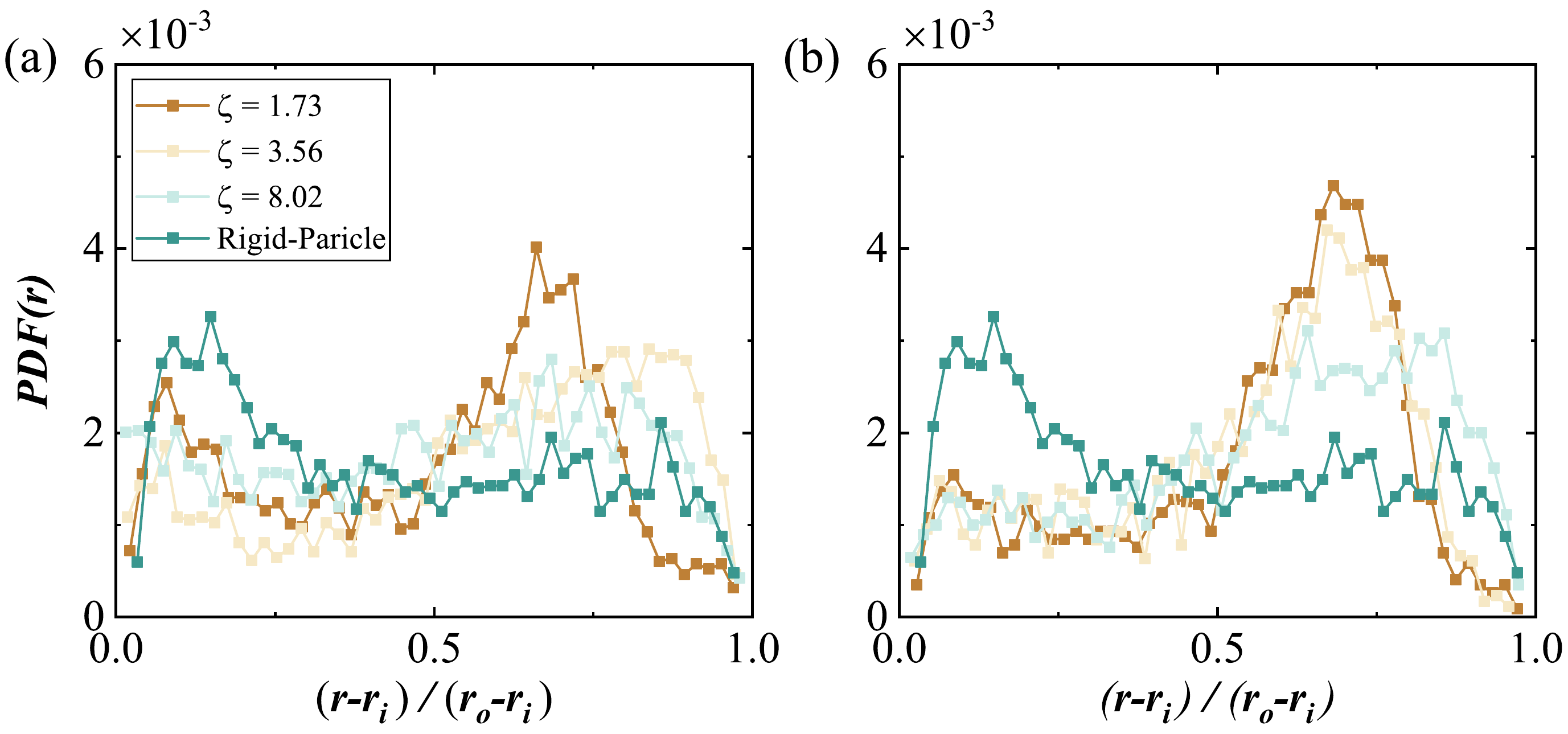}
	\vspace{-5 mm}
	\caption{ The droplet distribution along the radial direction of the system for (a) $\phi=2\%$ and (b) $\phi=4\%$ at $\dot{\gamma}=125$ s$^{-1}$. The distribution of rigid spherical particles taken from \cite{wang2022finite} is depicted for comparison.} 
	\label{fig8} 
\end{figure*}

It has been found that the magnitude of drag modification is connected to the preferential clustering in rigid particle suspensions \citep{calzavarini2008quantifying, wang2022finite} and bubbly flows \citep{colin2012turbulent, van2013importance, maryami2014bubbly, almeras2017experimental, mathai2018dispersion}. Using the camera mounted on the side of the system, the droplet distribution along the radial direction of the system can be obtained and analyzed by analogy with the rigid particle suspension. Since the droplet volume fractions are high, the droplet locations are manually detected at the smallest Re (or $\dot{\gamma}$), where the droplet can be well detected as mentioned before. The results are shown in figure~\ref{fig8}. Though the droplets are almost spherical (see figure~\ref{fig1}b), they show a preferential clustering in the bulk region, which has been reported in previous studies \citep{hudson2003wall, roccon2017viscosity, crialesi2022modulation} and can be attributed to the combined effect of wall migration and shear-induced diffusion of droplets. 

Compared to the rigid spherical particles, the droplets distribute even more towards the centre of the system and are in line with the results in previous numerical studies \citep{de2019effect,cannon2021effect}, where they report a migration towards the wall by prohibiting the droplet coalescence by applying a collision force. While as the $\zeta$ increases, the droplet distribution migrates towards the walls as the coalescence effect is increasingly suppressed. Instead of near the walls, the droplets show stronger clustering in the bulk than the rigid particles, partly being the reason that the $c_{f,\phi}/c_{f,\phi=0}$ of the emulsion flow is smaller than that of the suspension of rigid particles since the larger drag is connected to the near-wall clustering \citep{wang2022finite}.

\section{Conclusion}
The effects of the droplet viscosity on the turbulent emulsion rheology (global transport quantity) and on the droplet size distribution are experimentally investigated in Taylor-Couette turbulent flow. 
Based on density matching, the current system is governed by three dimensionless parameters, i.e., the viscosity ratio between two phases ($0.53\leq\zeta\leq8.02$), the Reynolds number ($4.2\times10^3\leq Re\leq2.8\times10^4$), and the droplet volume fraction ($0\%\leq\phi\leq10\%$). The drag of the emulsion is quantified by the normalized friction coefficient, $c_{f,\phi}/c_{f,\phi=0}$, to emphasize the magnitude of the drag variation induced by dispersed droplets.

It is found that, within the current investigated parameter regime, the $c_{f,\phi}/c_{f,\phi=0}$ is always larger than unity. The fact that the dispersed droplets always result in an increment of the drag than the corresponding single-phase flow could be induced by the extra interfacial tension contribution of the droplets. For a fixed $\zeta$, the droplet emulsions show a weaker $Re$-dependence than the rigid particle suspensions reported by \cite{wang2022finite} due to its vanishing finite-size effect and being in a tracer-like regime. Given this, the effects of droplets on the system drag mainly depend on the total surface area of the dispersed phase and are thus proportional to the volume fraction of droplets. This could account for the positive $\phi$-dependence of the system drag. On the other hand, the $c_{f,\phi}/c_{f,\phi=0}$ with respect to $\zeta$ shows two different stages of evolution. At the first stage, the $c_{f,\phi}/c_{f,\phi=0}$ is positively associated with the $\zeta$, which, as evidenced by the analysis of DSDs, is attributed to the suppression of coalescence effect as the $\zeta$ increases, therefore resulting in the increment of the total surface area. However, when the $\zeta$ increases beyond a critical value (here $1.7\leq\zeta_c\leq3.8$), the $c_{f,\phi}/c_{f,\phi=0}$ appears to be saturated at a plateau value depending on the turbulent intensity, which is almost equal to that of rigid particle suspensions.

In order to provide insights into the physics of observed drag modification, we perform image analysis to access the DSDs and the droplet distribution along the radial direction of the system. The mean droplet diameter is found to be almost independent of the $\zeta$ and $\phi$, while the droplet size range increases with increasing $\zeta$. The droplet formation mechanism is discussed by comparing the experimental data of DSDs with the $d^{-3/2}$ and $d^{-10/3}$ scaling. With increasing $\zeta$, the $d^{-3/2}$ scaling is principally identified for droplets with small size, hinting that the interfacial tension might become strong enough to sustain the droplet stability. While for the large-size droplets, the DSDs are generally captured by the $d^{-10/3}$ scaling, which implies the coalescence could be neglected and the droplets may experience a purely inertial breakup process. In this work, it is found that the scale of the transition point of two scalings is smaller than the Hinze scale. This could be the result of the inhomogeneity and anisotropy of the TC turbulence since the droplets might be fragmented by the fluctuation within the boundary layer. In addition, the droplet distribution along the radial direction of the system at small $Re$ shows a bulk preferential clustering, accounting for the smaller drag amplification of emulsion than the rigid particle suspension. The droplet distribution migrates towards the wall with increasing $\zeta$, leading to the same conclusion as the analysis of DSDs that the coalescence effects are suppressed when the droplets are highly viscous. Currently the image analysis can be performed only at small $\dot{\gamma}$ (or $Re$), which shed limited lights on the cases of high $\dot{\gamma}$ (or $Re$). In addition, the information of the carrier flow fails to be measured due to the interference caused by the droplets of high volume-fractions. Further studies are needed to explore the highly turbulent regimes and the in-depth of carrier flow.\\

\noindent {\bf Acknowledgements:}
We thank Federico Toschi, Detlef Lohse, and Sander Huisman for insightful discussions over the years, and Longfei Wang for his assistance with the experiments. This work was supported by the Natural Science Foundation of China under grant no. 11988102 and 91852202, and the Tencent Foundation through the XPLORER PRIZE.\\

\noindent {\bf Declaration of interests:} The authors report no conflict of interest.\\

\end{document}